\documentclass{article}

\usepackage{amsfonts}
\usepackage{epsfig}
\begin{document}

\title{Nonlinear Diffusion Through Large Complex Networks
Containing Regular Subgraphs}

\vspace{1cm}

\author{ {D. Volchenkov} \footnote{The Alexander von Humboldt Research Fellow at the
BiBoS Research Center}  and {Ph. Blanchard}
\vspace{0.5cm}\\
{\it  BiBoS, University Bielefeld, Postfach 100131,}\\
{\it D-33501, Bielefeld, Germany} \\
{\it Phone: +49 (0)521 / 106-2972 } \\
{\it Fax: +49 (0)521 / 106-6455 } \\
{\it E-Mail: VOLCHENK@Physik.Uni-Bielefeld.DE}}

\date{\today}
\maketitle

\begin{abstract}
Transport through  generalized trees is considered. Trees contain
the simple nodes and supernodes, either  well-structured
regular subgraphs or those with many triangles. We observe a
superdiffusion for the highly connected nodes while it is Brownian
 for the rest of the nodes. Transport within a
supernode is affected by the finite size effects vanishing as
$N\to\infty.$ For the even dimensions of space, $d=2,4,6,\ldots$,
the finite size effects break down the perturbation theory at
small scales and can be regularized by using the heat-kernel expansion.

\end{abstract}
\vspace{0.5cm}

\leftline{\textbf{ PACS codes: }87.15.Vv, 89.75.Hc, 89.75.Fb }
\vspace{0.5cm}

\leftline{\textbf{ Keywords: }Structures and organization in
complex systems, Networks and genealogical trees, Diffusion}

\large

%\newpage

\section{Introduction}
\noindent

Complex networks have become a new paradigm in physics. They
have been studied extensively due to their relevance to many real
systems from the World Wide Web to the biological and social
networks. Graph theory has been successfully applied to a wide
range of different disciplines requiring a description of sets of
elements either connected or interacting pairwise. Geometry and
topology of underlying graphs have a deep influence on the physical
properties of complex networks. In view of that one is interested
in the properties of graphs which affect the dynamical
behavior in the models defined on them. On the other hand, the study
of large complex systems calls for the statistical methods, to
give an effective description of collective dynamical behavior
observed in them.

In the present paper, we study the nonlinear transport  on
large inhomogeneous graphs.

Diffusion processes defined on the 1-dimensional  small world
networks have been considered in \cite{M}. It appears that the mean
field theory breaks down for the small worlds models in dimensions
$d<2$ \cite{KHK} due to the emergence of strong site-to-site
fluctuations of the Green's functions. In \cite{H}, the
$\epsilon$-expansion for a small world model has been developed where
$\epsilon=2-d$ (quantifies the departure of the space dimension
from $2$). The small world network is constructed by adding
random shortcuts to a regular lattice, and  the
density of links $p>0$ has been taken enough small: $pa^d\ll 1$ where $a$
is the lattice scale. The Green's functions
have been computed for $d=2$, but the $\epsilon$-expansion breaks down
for $\epsilon=1$ due
to traps appearing in the system at $d=1$.
In contrast to the small worlds model \cite{H}, we consider the
diffusion processes defined on graphs
approaching the networks known in the literature as "froths" \cite{Aste};
they could contain the regular subgraphs but are, in general,
far from  being  the regular lattices.
The space tiled
by the froth can be curved, and, in such a case, the intrinsic dimension
of the cellular system does not coincide with the dimension of the
embedding space.

Transport phenomena had been extensively studied
for the disordered media \cite{LB}-\cite{VD} and fractals
\cite{BM}-\cite{CR}. It has been revealed that the transport
properties  depend
upon the intrinsic dimension $d_f$ of disordered
systems. In particular, the random walk probability $P$ is proportional to
$ t^{-d_f/2}f(x/\sqrt{t})$ where $f$ is a scaling function.
Similar results have been reported for nonlinear diffusion
processes \cite{GMOL}-\cite{AH}.

In our model, we consider the complex networks as the generalized
trees with two types of nodes: the simple nodes (more often of low
connectivity) and the supernodes which are either the subgraphs
containing many polygons or the $k$-regular subgraphs. In a continuous
setting,  the supernodes can be treated as complete compact curved
Riemann  surfaces characterized by the finite areas (see Sec.~3).
The supernodes are bridged by
the tree components. We study the large scale asymptotic behavior
of nonlinear transport process defined on such the
generalized trees.
It is affected by both  the space curvature within a supernode
and the variation of space dimension between the supernodes.

%%%%%%%%%%%%%%%%%%%%%%%%%%%%%%%%%%%%%%%%%%%%%%%%%

Strictly speaking, the possibility to replace a finite graph by a compact
continuous manifold for a nonlinear diffusion process is a challenging question.
 For the linear differential operators acting on the periodic functions, it requires
 that the spectra associated  to the discrete and continuous problems are similar
\cite{Colin98}. In particular, the multiplicities of the eigenvalues
should be equal for both problems.

However, by this time, it is still not much known even about the spectral properties
of Schr\"{o}dinger operators defined on  curved compact manifolds. Some estimations on the
 multiplicities of eigenvalues
and the Euler characteristic of a surface can be found
for the 2D-sphere, for the Klein's bottle, and for the 2D-torus in
 \cite{Cheng}-\cite{Colin87}.
Therefore, even in this case, the justification of  the validity for the
 approximation of processes defined on
the graphs by that ones defined on the compact continuous manifolds is a
difficult problem.
Concerning the nonlinear diffusion process, we should confess that such an
 approximation is always an assumption.

We consider the diffusion as a generalized Brownian motion with
 arbitrary boundary conditions described by a functional
integral (see Sec.~4). The nonlinear term included into the
diffusion equation models the effect of varying space dimension,
the possible fluctuations of transport coefficient, the diffusion-reaction
processes, and the possible queueing due to a bounded
transport capacity of edges bridging the supernodes (see Sec.~5).
The fluctuations have been treated in the framework of field
theory approach.

The large scale asymptotic behavior of Green function is a
superdiffusion for the nodes of high connectivity,  while it is
still Brownian for the rim nodes (see Sec.~6).
In the regular subgraphs endowed with the standard orientation of
edges, the effect of space "curvature" within a  supernode
rises the finite size corrections to the scaling behavior (see Sec.~6
and Appendix A). In Sec.~7, we use the heat-kernel expansion to regularize the
effective action functional  at the small scales for the even
dimensions of space  $d=2,4,6,\ldots.$

\section{Description  of the model and the results}

We are interested in the large scale asymptotic behavior of Green
function for the diffusion
equation for the density $u(x,t)$ defined
on the curved Riemannian manifold with  metric tensor $g_{ij}$,
with the nonlinear term $\propto u^\alpha$ ($\alpha>1$) included.
We assume the such a model can be considered as an approximation of
nonlinear diffusion process defined on the graphs containing the large
regular subgraphs. This approximation, however, is rather intuitive and
cannot be proven nowadays rigorously. If the assumption is true, then
$\log_2\mathrm{deg}(x)$ plays the role of effective local space dimension
 at node $x$ on the graph, $\mathrm{deg}(x)$ being the degree of $x$.
The nonlinear diffusion
  term is relevant to the large scale asymptotic behavior if
$$   \alpha \leq 1+\frac 2{\log_2 \textrm{deg}(x)}  $$
and is irrelevant otherwise.

In the critical phenomena theory, the physical degrees of freedom are
 replaced by the scaling ones related to each other by the RG
 transformations. The large scale asymptotic behavior of physical system
 is then determined by
 the properties of scaling invariant system at a certain stable fixed
  point of the RG differential equation. The derived results are irrelevant
  to the order of space and time (infinite) limits.

Our main result is that for the large scale asymptote of Green function,
\[
G(t)\sim \left\{
\begin{array}{ll}
  t^{-\log_2 \mathrm{deg}(x)/2}, & \log_2\mathrm{deg}(x)>\frac 2{\alpha-1},
  \quad x\in \Gamma, \\
  t^{-1/\alpha-1}, &\log_2\mathrm{deg}(x)\leq \frac 2{\alpha-1},
  \quad x\in \Gamma,
\end{array}
\right.
\]
The main technical difficulty of computations in the curved space metrics
is the lack of the global frequency-momentum representation.
In order to
regularize the effective action at small scales in the one-loop order,
we have used
the heat-kernel expansion which does not depend upon the space-time topology.
In the regular graphs endowed with the standard orientation of
edges (the cycling ordering of edges to be the same for all
nodes), the effect of curvature within the supernodes reveals
itself by the finite size corrections to the scaling behavior.

%%%%%%%%%%%%%%%%%%%%%%%%%%%%%%%%%%%%%%%%%%%%%%%%%

\section{The regular subgraphs viewed as Riemann surfaces}
\noindent

In the present paper, we consider the flat Euclidean space $\mathbb{R}^d$
 as the limit $a\to 0$ of a regular lattice
$\mathcal{L}_a=a\mathbb{Z}^d$ with lattice scale length $a$.
For simulating the diffusion equation $u_t = \triangle u$ for
the scalar function $u$ defined on the lattice, one uses its
discrete representation,
\begin{equation}\label{00}
  u^{t+1}(x) = \frac{1}{2^d a^2}\left[ \sum_{y\in U_x} u^t(y)
  - 2^d u^t(x)\right],
\end{equation}
where  $U_x$ is the lattice neighborhood of $x\in \mathcal{L}_a$.
 The cardinal number $2^d$ is uniform for a given
 $\mathcal{L}_a$, and $d$ is interpreted as the dimension of Euclidean space.

Being defined on an arbitrary connected graph
$\Gamma=(V_\Gamma,E_\Gamma),$ in which $V_\Gamma$ is the set of
its vertices and $E_\Gamma$ is the set of edges linking them, the
discrete Laplace operator has actually the same form as in
(\ref{00}) excepting for the cardinality number changed to
$2^{\delta_x}$ where $\delta_x = \log_2 \mathrm{deg}(x)$ and
$\mathrm{deg}(x)$ is the degree of vertex $x\in\Gamma$. By
counting the number of vertices $|V_\Gamma|$, the number of edges
$|E_\Gamma|$, and the number of faces $|F_\Gamma|$, one obtains
the Euler characteristic of a planar graph $\Gamma$,
\[\chi(\Gamma)
=|V_\Gamma|- |E_\Gamma|+ |F_\Gamma|,
\]
and its genus,
\begin{equation}\label{genus}
  \mathrm{g}(\Gamma)=\frac{2-\chi(\Gamma)}{2},
  \end{equation}
 the  number of nonintersecting cycles on the graph $\Gamma$.
We are interested in the subgraphs of $\Gamma$ ample with
polygons, in particular, the triangles, which can be used as a base
for a local mesh of a Riemann surface (the Delaunay
triangulation).

 An intuition on the deep relation between the
triangle structures in the complex networks and their curvature
has been expressed in \cite{EM}: "Triangles capture transitivity,
which we measure by the associated notion of curvature". In
\cite{CE},  upper and lower bounds on the number of graphs of
fixed degree which have a positive density of triangles have been
estimated. In particular, it has been shown that the triangles
seem to cluster even at low density. Then, in \cite{Sergi} it has
been proven that the probability for a randomly selected vertex to
participate in $T$ triangles decays  with $T$ following a
power-law, if the graph is scale free with the degree
exponent satisfying $2<\beta<2.5$. Moreover, if $\beta=2+1/3$,
the density of
triangles appears to be finite  in such graphs.

 The Laplace-Beltrami operator on a triangle mesh has been
 defined in connection with the various graphical applications
 such as mesh fairing, smoothing, surface editing in $3D$-space
 \cite{KKDS}-\cite{KR}. The stiffness matrix correspondent to
 it can be computed for each
triangle specified by its vertices $\mathbf{x}=(x_1,x_2,x_3)$,
$\mathbf{y}=(y_1,y_2,y_3)$, $\mathbf{z}=(z_1,z_2,z_3)$ as
\[
\mathbf{K}=S_\triangle\mathbf{B}\mathbf{B}^T, \quad \mathbf{B}=
\frac{1}{2S_\triangle}\left(
\begin{array}{ccc}
   x_2-x_3 & y_2-y_3 & z_2-z_3  \\
  x_3-x_1 & y_3-y_1  & z_3-z_1  \\
  x_1-x_2 & y_1-y_2  & z_1-z_2
\end{array}
\right)
\]
where $S_\triangle$ is the area of triangle.

However,  from the various models of complex networks, it seems
rather easier to count the number of edges which link a node to
others than to check out if its neighbors are really connected
forming  triangles. Instead of hunting for triangles, while
analyzing the graphs of real world networks, one can search for
the 3-regular subgraphs $_{3\mathrm{R}}\Gamma$ for which there
are precisely three edges incident at each vertex (loops are
counted twice, multiple edges are allowed). These graphs  form a
honeycomb. The idea of using 3-regular subgraphs to study the
topological properties of rendered Riemann surfaces has been proposed in
\cite{Bu}. In \cite{BMa, Man}, it has been shown that for each
3-regular graph ${}_{3R}\Gamma$ of $2N$ nodes  with an orientation
$\mathcal{O}$ (which assigns to each vertex of ${}_{3R}\Gamma$ a
cyclic ordering of edges incident at it) one can construct a
complete Riemann surface $S({}_{3R}\Gamma,\mathcal{O})$ by
associating the ideal hyperbolic triangles to each vertex of
${}_{3R}\Gamma$ and gluing sides together according to the edges
of the graph ${}_{3R}\Gamma$ and the orientation $\mathcal{O}.$
The resulting surface is endowed with a  metric excepting for the
finitely many points (cusps) where the metric could be undefined.
The surface area is finite and equals to $2\pi N$. The conformal
compactification of $S({}_{3R}\Gamma,\mathcal{O})$ are dense in
the space of Riemann surfaces. The standard orientation on the
3-regular graph which is the 1-skeleton of a cube contains six
left-hand-turn paths on which a traveller always turns left,
giving that the associated surface is a sphere with six punctures.
The choice of different orientations can give  surfaces of
genus 0,1, and 2, \cite{BMa}.

The closed paths of length $k$ on $_{3\mathrm{R}}\Gamma$ are
homotopic to the closed geodesic lines on
$S(\Gamma,\mathcal{O})$. Their cardinality is defined by the
spectral density of $_{3\mathrm{R}}\Gamma$-subgraph that is the
density  of eigenvalues of its adjacency matrix \cite{FDBV},
\[
\varrho(\lambda)=\frac{1}{N}\sum_{j=1}\delta(\lambda-\lambda_j),
\]
which converges to a continuous function with $N\to\infty$
($\lambda_j$ is the $j-$th largest eigenvalue of adjacency matrix
$\mathbf{A}$).  The $k-$th moment $M_k$ of $\varrho(\lambda)$
scales with the number of cycles $C_k$ of length $k$,
\[
M_k=\frac{1}{N}\sum_{j=1}^{N}\lambda_j^k=\frac{1}{N}\mathrm{Tr}(\mathbf{A}^k)
=\frac{k}{N}C_k,
\]
 $C_k$ can be computed by using of the zeta-function,
\[
\zeta(z)=\frac{1}{\det(\mathbf{1}-z\mathbf{A})}.
\]
Namely,
\[
C_k=\left.\frac{1}{(k-1)!}\frac{d^k}{dz^k}\zeta(z)\right|_{z=0}.
\]
All roots of the polynomial $\det (\mathbf{1}-z\mathbf{A})$ lay on
the unit circle and coincide with the inverse of the eigenvalues
$\lambda_j^{-1}$ of adjacency matrix excepting for the zeros,
$\{\lambda=0\}$. It is worth to mention that
$S({}_{3R}\Gamma,\mathcal{O})$ allows for a conformal
compactification, \cite{BMa}.

\section{Model of nonlinear diffusion through  complex networks}
\noindent

In  the previous section, we have supposed that a regular subgraph
in a complex network can be treated, in a continuous setting, as a
Riemann space of finite area characterized by some metric tensor.
For a particular time slice, the line element $ds$ between each
pair of neighboring points on the spatial surface is given by
\[
ds^2=\sum_{ij}g_{ij}(x)dx^idx^j
\]
where $dx^k$ denote the  differences between
neighboring points, and $g_{ij}$ denotes the metric tensor. Then,
the complex network as a whole can be considered as a disordered
media in which the compact islands $S$ of curved Riemann space
(with the "effective" space dimensions $\delta_S$) are bridged by
the tree like graph components (see Fig.~1) in which the local
space dimension $\delta_x$ can vary from point to point.

The transport properties through such a disordered media is essentially of
 nonlinear nature. In the previous studies of nonlinear
diffusion \cite{GMOL} -\cite{AH}, the authors had introduced
various nonlinear terms into the diffusion equation modelling the
possible fluctuations of transport coefficient, the
diffusion-reaction processes, and the queuing due to a bounded
transport capacity of edges. We also introduce it for accounting
the effect of varying dimension of space in the complex network
(see the discussion below). To be certain, let us consider the
equation for the scalar density field $u(\mathbf{x},t)$ defined on a
Riemann surface,
\begin{equation}\label{eq1}
  \nabla_{t}u =g_0^{ij}\nu_0\frac{\partial^2 u}{\partial x_i\partial
  x_j}-\xi_0\nu_0 R u - \eta_0 \nu_0 u^\alpha, \quad
  \nabla_{t} \equiv \partial_{t}+ b_0^i\frac{\partial}{\partial x_i}.
\end{equation}
All $0$-subscripted variables denote their bare values before the
application of the renormalization group transformation. In
Eq.~(\ref{eq1}), the Riemann metric tensor $g_0^{ij}$ depends upon
the chosen conformal parameterization of regular subgraphs, $R$ is the scalar curvature. We
prefer to keep the entries of $g_0^{ij}$ dimensionless therefore
we have introduced the parameter $\nu_0$ having the
dimension of a viscosity. For the purpose of this paper, we will
examine metric rescalings that are spacetime constants (we suppose
that the subgraph of $\Gamma$ is regular and the edges do not
rewire with time). However, it is possible to consider the effect
of a rescaling given by a spacetime-dependent function.

The covariant derivative $\nabla_{t}$ contains the curvature drift
term proportional to $b_0^i=g_0^{jk}\Gamma^i_{jk}$ (the curvature
drift velocity) which expresses the local anisotropy of space
because of its curvature. $\Gamma^i_{jk}$ are the Christoffel
symbols calculated out from the metric tensor $g^{ij}$ in the
standard way,
$\Gamma^i_{jk}=g^{il}(g_{kl,j}+g_{jl,k}-g_{jk,l})/2$. While being
interested in the long time large scale ranges, one usually keeps
only the first order derivative term $\partial_i$ if it presents
in the linear part of equation since its contribution $O(k)$
should dominate over the diffusion $O(k^2)$ for small $k$.
Nevertheless, we keep both terms to ensure the convergence of
integrals in time, in the limit of flat metric. However, in the
general curved spacetime, the homogeneity required for the
existence of a global momentum space representation is lacking.

The real valued parameters  $\xi_0$  and $\eta_0$ are the bare
coupling constants governing the coupling of configurations $u$ to
the scalar curvature of space $R$ and to the varying effective
space dimensionality $\delta_x$ respectively. We assume them to be
small. The nonlinearity exponent is $\alpha>1$ ($\alpha$ is not
necessary integer) \cite{GMOL} -\cite{AH}.

%%%%%%%%%%%%%%%%%%%%%%%%%%%%%%%%%%%%%%%%%%%%%%%%%%%%%

Let us explain the role played by  the nonlinear term in (\ref{eq1}) in
more details.  In the critical phenomena theory, the physical degrees of
 freedom are replaced by the scaling degrees of freedom. In particular,
  one considers the canonical dimension $d_F$ instead of time and space
  physical dimensions of the quantity $F$. The analysis of canonical dimensions
   allows for the selection   of relevant interactions among all possible
   interaction which could arise in the model. In the spirit of critical
    phenomena theory, the nonlinear term that could affects substantially
    the large scale asymptotic behavior typical for the diffusion process
     should have the same canonical dimension as the normal diffusion. If
     the canonical dimension of the nonlinear term added to the equation
     is less than of the ordinary diffusion, it has to be neglected. The
     opposite is also true: if the nonlinear term provides the leading
      contribution to the asymptotic behavior, then the diffusion term
      has to be dropped.

Therefore, it is interesting to consider the model in which the nonlinear term
would play an important role. Below, we demonstrate that the
exponent $\alpha$ is related to the dimension of space $d$ and the dimension
$[\eta_0]$ of coupling constant $\eta_0$.  If we assume for a moment that in
 the vicinity of some point the dimension of space $d$ is changed to some
 other value $\delta$, then, strictly speaking, Eq.~(3) could have no sense
 therein: either the diffusion term or the nonlinearity should be neglected.
 For given $\alpha$ and $[\eta_0]$,  there is only one value $d$ at which
 the Eq.~(3)  is relevant with respect to the large scale asymptotic behavior
 of diffusion process.
  If we consider the plane of parameters
 $[\eta_0]$ and $\alpha$, then the relevant space dimensions $d$ is a line
 on it. Therefore, by tuning the values of $\alpha$ and $[\eta_0]$, one can
  "modify" the space dimension $d$ in the model of nonlinear diffusion. It
  is indeed unphysical to change the nonlinearity exponent $\alpha$ (we suppose
  that $\alpha$ is a property of a certain physical process), however, one can
   tune the value of $[\eta_0]$ and use it as the small expansion parameter
    of perturbation theory (like the parameter $\varepsilon=4-d$  in the Wilson's
theory  of   critical phenomena).

A similar idea is used in the usual dimensional regularization of Feynman
 diagrams. In the continuous Euclidean space,
 the dimensional regularization scheme does not look natural and therefore
  is usually treated as a formal trick which helps to reformulate the
  singularities arisen in the Feynman graphs in the form of poles in
   $\varepsilon$.  However, if the dimension of physical space could
   vary, than the dimensional regularization would acquire the natural
   meaning provided the nonlinearity exponent $\alpha$ is fixed and the
   correspondent nonlinear diffusion process is relevant with respect to the
 large scale asymptotic behavior.

%%%%%%%%%%%%%%%%%%%%%%%%%%%%%%%%%%%%%%%%%%%%%%%%%%%%

 As we have mentioned above, such a relevance  can be justified by means of dimensional
analysis \cite{BK,AH}. Dynamical models have two scales: the length scale $L$ and the
time scale $T$. The physical dimension of viscosity  is
$[\nu_0]=L^2T^{-1},$ of the scalar curvature is $[R]=L^{-2}$, and of the drift velocity is
 $[b_0^i]=LT^{-1}$. Let us choose the physical dimension for the
coupling constant to be $[\eta_0]=L^{-2\varepsilon}$ (assuming
that $\varepsilon=0$ in the logarithmic theory when the nonlinear
interaction is marginal) and note that the dimension of a scalar
quantity is $[u]=L^{-d}.$ In the free theory, $\partial_t \propto
\partial^2$ and in the analysis of canonical dimensions one has
 $L^2\sim T$. All terms in (\ref{eq1}) should be of the same
canonical dimension, in particular $[\partial_t u]=[\eta_0\nu_0
u^\alpha ]$, therefore $-2+\log_L[u]=\alpha \log_L[u]-2\varepsilon $ and
finally, $[u]=L^{2(1-\varepsilon)/(1-\alpha)}$, from which it follows that
\begin{equation}\label{d}
  d = 2\cdot\frac{1-\varepsilon}{1-\alpha}.
\end{equation}
The above relation gives us a hint of that the space
dimension in the model of a nonlinear diffusion can be
effectively  tuned by the parameters $\alpha$ and $\varepsilon.$
We choose the parameter $\varepsilon$ to quantify the local
irregularity of the graph by measuring the relative deviation of the
node degree  $2^{\delta_x}$ from the cardinality number $2^d$ in
the regular lattice,
\begin{equation}\label{eps}
  \varepsilon =1 -\frac{d}{\delta_x}.
\end{equation}
The nodes with $\mathrm{deg} (x) < 2^d$
correspond to $\varepsilon <0$, while the nodes for which
$\mathrm{deg} (x) \geq 2^d$ are described by $\varepsilon \geq 0.$

We supply Eq.~(\ref{eq1}) with the locally integrable initial
condition $u(\mathbf{x},0)$ and study the standard Cauchy problem
being interested in the large scale asymptotic Green's
functions $G(\mathbf{x},\mathbf{x_0};t,t_0).$ For a curved space,
the natural way to proceed is to examine the change to the Green's
functions as the metric is scaled. This can be achieved by moving
the points along the geodesics (cycles in the graph $\Gamma$)
connecting them or alternatively by scaling the geodesic distance
function (the metric).

The Green's functions of nonlinear problem (\ref{eq1}) supplied
with the integrable initial conditions can be formally calculated
by the perturbation series with respect to the nonlinearity (as
the coupling parameter $\eta_0$ is small) followed by the
integrations over the initial condition $u(\mathbf{x},0).$ Some
integrals estimating corrections to the linearized diffusion
problem diverge logarithmically since the integration domain is
not compact. If we introduce the $\varepsilon$-parameter in
accordance to (\ref{eps}), the divergences reveal themselves by
the poles in
$$
\varepsilon =1 + d\cdot \frac{1-\alpha}{2}.
$$
Therefore, the nonlinear interaction is irrelevant (in the sense
of Wilson) for $\varepsilon <0$ ($d>2/(\alpha-1)$), but is
essential as $\varepsilon \geq 0$ when the ordinary perturbation
expansion (in the form of series in $\eta_0$) fails to give the
correct large scale asymptotic behavior and the whole
series has to be summed up. For instance, it happens at $d=2$ for
$\alpha=2$.

In other words, the logarithmic (marginal) value of
$\alpha$ is determined by comparison of the nonlinear contribution
with that of the linear dissipative term, $\alpha{\ }_{\mathrm{log}}=
1+ 2/d$. While introducing the parameter $\varepsilon$ accounting
for the local change of connectivity in the graph, we effectively
pass from $\alpha{\ }_{\mathrm{log}}$ to its new value $\alpha{\
}'_{\mathrm{log}}= 1+ 2/\delta_x$. Then it turns that the
nonlinear contribution to the long range asymptotic transport
through the rims ($\varepsilon <0$) is irrelevant in comparison
with the linear diffusion and therefore can be neglected. In
contrast to it, the contribution coming from hubs ($\varepsilon
\geq 0$) is more essential than the linear one and has to be taken
into account in all orders of perturbation theory since the
relevant fluctuations dominate the diffusion at large scales.

We calculate the asymptotic Green's functions for the model
(\ref{eq1}) in the logarithmic theory (on the regular subgraphs
with the cardinality number $2^d$) in curved space metric and
develop the $\varepsilon-$expansion accounting for the corrections
in the long time large scale region due to the irregularity of
graph. The relevant contributions to the nonlinear transport
coming from hubs are summed by the field-theoretic renormalization
group method. Herewith, the real values of parameter $\varepsilon$
has to be taken as $ \varepsilon_x=1-{d}/{\delta_x},$ the excess
of hub's connectivity over the regular cardinality number $2^d$.

We conclude this section by remarking that the problem of
renormalization in a curved spacetime has been discussed
extensively in the literature (\cite{DeWitt}-\cite{T83} and by
other authors). However, it has never been studied in connection with
the critical phenomena theory. The analysis of transport through
the graphs would provide us with such a model. It is worth to
mention that in contrast to the case of the gravitational field,
 we are not restricted on
graphs by the equivalence principle, so that
the curved space has not to be flattened in any sufficiently small
region.

\section{Diffusion as a generalized Brownian motion}
\noindent

We set up a field theory formalism for the study of
asymptotic properties of the Green's function by means of
the renormalization group (RG) equations which are valid in the curved
spaces. Our method is closely related to the approach discussed in
\cite{NelPan} regarding the Green's functions as the functions of
 metric and scaling the metric instead of scaling the
coordinates or moments.

It is well known that many problems of stochastic dynamics (and of
the transport through a disordered media, in particular) can be
treated as a generalized Brownian motion, $P(u)={\left\langle
{\delta \left( {u - u\left( {\mathbf{x},t} \right)} \right)}
\right\rangle},$ in which the average is taken over all
configurations of field $u(\mathbf{x},t)$ satisfying the dynamical
equation
\begin{equation}\label{eq2}
  \nabla_t u - \nu_0\Delta_{LB} u + \nu_0\eta_0 u^{\alpha}+\xi_0\nu_0 R u= \frac 1{\sqrt{g}}
    \delta(t-t_0)\delta(\mathbf{x}-\mathbf{x}_0),
\end{equation}
for the integrable initial condition $u(\mathbf{x},0)$ and
\textit{arbitrary} conditions on the boundaries of the graph. The
Laplace-Beltrami operator $\Delta_{LB}$ is given by (\ref{eq1}),
and $g=|\det g_0^{ij}|.$

We use the functional representation of the  $\delta-$function for expressing the probability,
\begin{equation}\label{P}
  P(u)=\int \mathcal{D}u\int\mathcal{D}u'  \exp
\left(u'\left(u-u(\mathbf{x},t)\right)\right),
\end{equation}
in which $u$ marks the position of a "particle", and the auxiliary
field $u'$ (of the same nature as $u$) is not inherent to the
original model, but appears since we treat the dynamics as a
Brownian motion. The formal convergence requires the field $u$ to
be real and the field ${u}'$ to be purely imaginary.  Should a
unique  solution of dynamic equation exists, we perform the
natural change of variables in (\ref{P}),
\[
\left(u-u(\mathbf{x},t)\right) \to -\nabla_t u + \nu_0\Delta_{LB}
u - \nu_0\eta_0 u^{\alpha}-\xi_0\nu_0 R u- \frac 1{\sqrt{g}}
    \delta(t-t_0)\delta(\mathbf{x}-\mathbf{x}_0) =0,
\]
from which it follows that
\[
 P(u)=\int \mathcal{D}u \int\mathcal{D}u' \exp
\mathcal{S}(u,u')\det \mathbf{M}
\]
where $\det \mathbf{M}$ is the Jacobian associated to the change of
variable, and $\mathcal{S}(u,u'),$  the action functional,
\begin{equation}\label{S}
\mathcal{S}=\mathcal{S}_0 - \eta_0\nu_0 \mathrm{Tr}_g (u'u^\alpha)
-\frac{1}{\sqrt{g}}u'(\mathbf{x}_0,t_0), \quad \mathcal{S}_0=
\mathrm{Tr}_g\left( -u'\nabla_t u + \nu_0u'\Delta_{LB}
u-\xi_0\nu_0 R u'u\right).
\end{equation}
The trace  $\mathrm{Tr}_g  $ means  the summation over
the discrete indices and the integration $\int dv_x\int dt$ over
the invariant volume element on the $d$-dimensional manifold,
$dv_x=\sqrt{g(x)} d^dx.$

The Jacobian $\det \mathbf{M}$ deserves a thorough consideration.
The linear part of the variable transformation can be factorized from it,
\begin{equation}\label{detM}
\det \mathbf{M}=\det\mathbf{M}_0\det (1-\Delta_{uu'}\mathbf{M}_1)
\end{equation}
where $\mathbf{M}_0 = -\nabla_t + \nu_0\Delta_{LB}-\xi_0\nu_0 R $,
the interaction part
$\mathbf{M}_1=\alpha\nu_0\eta_0u^{\alpha-1}\delta(t-t')$, and
$\Delta_{uu'}$ is the Feynman propagator in curved Riemann space,
\cite{BuPa, Bala}, defined as the solution of linearized problem
\[
(-\nabla_t + \nu_0\Delta_{LB}-\xi_0\nu_0 R
)\Delta_{uu'}(\mathbf{x}-\mathbf{x}',t-t')=
\frac{1}{\sqrt{g}}\delta(t-t')\delta(\mathbf{x}-\mathbf{x}').
\]
For more details as well as its explicit form see
Appendix A.

It is important for us that the propagator $\Delta_{uu'}$ is
proportional to the Heaviside function $\theta(t-t')$ as a
consequence of causality principle. The first factor in
(\ref{detM}) does not depend upon fields and therefore can be
scaled out of the functional integration. The second factor in
(\ref{detM}) can be expanded into the "diagram" series,
\[
\log \det (1-\Delta_{uu'}\mathbf{M}_1)= -\mathrm{Tr}_g
(\Delta_{uu'}\mathbf{M}_1+\frac 12
\Delta_{uu'}\mathbf{M}_1\Delta_{uu'}\mathbf{M}_1+\ldots)
\]
comprising of cycles of retarded Feynman propagators proportional to the
Heaviside functions and therefore being trivial, excepting for the
very first term, $\mathrm{Tr}_g(\Delta_{uu'}\mathbf{M}_1)$, in
which the operator $\mathbf{M}_1$ contains $\delta(t-t')$. The
first term in the expansion is proportional to the undefined
quantity $\theta(0)$ which value is usually taken as $1/2$.
However, in the critical dynamics, another convention is used
\cite{Adzhemyan:1999}, $\theta(0)=0$ , under which the Jacobian
$\det \mathbf{M}$ turns to be just a constant and therefore may be
scaled out away at the irrelevant cost of changing only the
normalization.

The action functional of type (\ref{S}) in the problem of
nonlinear diffusion in the flat metric has been introduced in
\cite{AH}. In \cite{VoL}, the functional with an ultra-local
interaction term like in (\ref{S}) has been derived  as a limiting
one in the framework of  MSR formalism (stochastic quantization,
\cite{Martin:1976}). The renormalization of field theoretic models
with ultralocal terms, located on surfaces, had been studied in
\cite{Sym} in details.

Further insight into the field theory representation of Brownian
motion and the properties of auxiliary field $u'$ can be obtained
from the equations for the saddle-point configurations. The first
equation, $\delta\mathcal{S}/\delta u' =0,$ recovers the original
Cauchy problem. The other one, $\delta\mathcal{S}/\delta u =0,$
reads as following
\[
\nabla_t u'+\nu_0\Delta_{LB}u'-\xi_0\nu_0 R=\alpha\eta_0\nu_0
u'u^{\alpha}
\]
and is characterized by a negative viscosity. One can conclude
from it that the auxiliary field should be trivial for positive
time,  $u'(t>0)=0$, and decays as $t\to -\infty$.

In the framework of field theory approach, the Green function
$G(x,t;x_{0} ,t_{0} )$ for the Cauchy problem (\ref{eq2}) can be
computed as the functional average,
\begin{equation}\label{Green}
G(x,t;x_{0} ,t_{0} ) = {\left\langle {u} \right\rangle}  =
{\frac{{\int\mathcal{D}u \int \mathcal{D}u'{u(x,t)\exp S(u,u')}}
}{{\int\mathcal{D}u \int \mathcal{D}u' {\exp \left( {S_{0}}
\right)}} }},
\end{equation}
in which $\mathcal{S}$ is the action functional (\ref{S}). The
Green function (\ref{Green}) and all higher moments of fields $u$
and $u'$ allow for the standard Feynman diagram series expansions.
The diagram technique with the ultralocal interaction terms has
been discussed in \cite{AH}. A special feature of such diagrams is
that the final point of any diagram corresponds to
$(\mathbf{x}_0,t_0)$. It is worth to mention that diagrams could
formally contain a non-integer number of lines (since the
nonlinearity exponent could deviate from an integer number).
Diagrams are drawn of three elements: i) the
final point $(\mathbf{x}_0,t_0)$ with an arbitrary number of
attached $u'$-legs (we mark them by a slash, see Fig.2); ii) the
interaction vertex with one $u'$-leg and $\alpha$ $u$-legs
attached to it (we put the letter $\alpha$ inside the loop to
stress that it is not necessarily integer); iii) the propagator
$\Delta_{uu'}$ is only available. The first three diagrams
for the Green function (\ref{Green}) are sketched in Fig.~2.

\section{Renormalization Group equation and scaling behavior of
scalar field coupled to a complex network}
\noindent

Power-counting arguments \cite{AH} show that the divergent Feynman
diagrams are those which involve any number of external $u'$-legs
(lines with slashes). For all these functions the formal index of
divergence equals to zero, therefore all divergent contributions
are logarithmic (the correspondent counterterms are constants). It
is worth to mention that the model is renormalizable despite the
fact that it requires an infinite number of counterterms. It is
sufficient to renormalize the only "one-particle-irreducible"
Green function (the only diagram block which can be drawn using
the elements i) - iii) mentioned above) to render the model
finite.

Moreover, the counterterm corresponding to this block is sufficient
to regularize all higher moments of fields $u$ and $u'$, since any
diagram of perturbation series is expressed as a convolution of
equivalent blocks. Diagrams contain no additional superficial
divergences \cite{AH}. We recall that all loops which could arise
in the diagram expansions are created by the single local vortex
with any number of $u'-$legs incident at it. The counterterm
corresponding to the elementary divergent block is constant and
local in configuration space, i.e. $\propto
\delta(t-t_0)\delta(\mathbf{x}-\mathbf{x}_0)$. In the action
functional (\ref{S}), the same local term appears, so that the
model is renormalized multiplicatively, and only
the renormalization constant $Z$ is required. To keep the interaction
coupling constant $\eta$ dimensionless, one must introduce a mass
parameter $\mu$.

The bare parameters are related to the renormalized parameters by
\[
\nu_0=\nu, \quad \xi=\xi_0,
\]
\begin{equation}\label{000}
 \quad \eta_0=\eta\mu^{2\varepsilon}Z^{\alpha-1}.
\end{equation}
 The auxiliary
fields and the Green function are related to their renormalized
analogs by:
\[
u'=u'_RZ,  \quad
G(\nu_0,\xi_0,\eta_0)=Z^{-1}G_R(\nu,\xi,\eta,\mu).
\]
The only renormalization constant required in the model is

\begin{equation}\label{Z}
  Z=1 + \sum_{l=1} \frac{c_l(\xi,\eta,\mu)}{\varepsilon^l}
\end{equation}
where the amplitudes $c_l$ are defined to be precisely those
needed to subtract the poles in the corresponding Feynman
integrals. The derivation of the RG equation is carried through in
a standard way since the curvature coupling parameter $\xi$ is not
renormalized. We apply the standard RG differential operator
$\mu\partial_\mu$ (for fixed $\nu_0$, $\eta_0$, $\xi_0$) to $G_R$
to obtain
\begin{equation}
\left[ \mu\frac{\partial}{\partial \mu}+ \mu\frac{\partial
\eta}{\partial \mu}\frac{\partial }{\partial \eta}-\mu
\frac{\partial }{\partial \mu}\ln Z
\right]G_R(\nu,\mu,\xi,\eta)=0.
\label{RGEq}
\end{equation}
It follows from the relation (\ref{000}) that for the RG functions
(the coefficients in the RG equation) one has
\begin{equation}\label{ffg}
  \gamma(\eta)=\mu \frac{\partial}{\partial\mu} \ln Z, \quad
\beta(\eta)=\eta\left(-2\varepsilon-(\alpha-1)\gamma(\eta)\right).
\end{equation}
In the forthcoming section, we discuss the existence of a fixed
point of RG transformation stable with  respect to the long time
large scale asymptotic behavior, that is a solution  $\eta_*$ of
equation $\beta(\eta)=0$ such that $\beta '(\eta_*)>0$ in the
physical range of parameter $\eta>0$.

In the equation of critical scaling for the curved space, one has
to take into account the new mass parameter  $\varpi$ in addition
to $\eta,$ $\xi$, and $\mu$. It had been introduced in
\cite{NelPan} to quantify the dilatation  $\varpi^{-1}g_{ij}$ of
the background metric  $g_{ij}$. The derivation of the scaling
equation is standard: the canonical scale invariance of the
renormalized Green function $G_R(\xi,\mu,\nu,\eta)$ with respect
to dilatations of all variables is expressed by the equations
\begin{equation}\label{sc}
  \left[\mu\frac{\partial}{\partial\mu}+
  \varpi\frac{\varpi}{\partial \varpi} -
  2\nu\frac{\partial}{\partial \nu}
  -x\frac{\partial}{\partial x} - d\right]G_R(\xi,\mu,\nu,\eta)=0,
\end{equation}
\[
\left[ -t\frac{\partial}{\partial t} +\nu\frac{\partial}{\partial
\nu} \right]G_R(\xi,\mu,\nu,\eta)=0.
\]
In the IR fixed point of RG-transformation, one  excludes the differential
operations $\mu \partial_ \mu$ and $\nu \partial_ \nu$  using
 the RG equation (\ref{RGEq})
to obtain the equation of critical scaling,
\begin{equation}\label{sceq}
 \left[ -\varpi\frac{\varpi}{\partial \varpi}+d - \gamma
\right]G_R(\xi,\mu,\nu,\eta)=0,
\end{equation}
where $\varpi$ is now interpreted as the metric scaling variable.
If a stable point of
RG transformation  relevant to the long range asymptotical
behavior $\eta_*$ exists in the model, the value of the anomalous
dimension $\gamma$ at the fixed point is found exactly, owing to
the relation between $\beta$ and $\gamma$,
\begin{equation}\label{gan}
 \gamma_*=\gamma(\eta_*)=-\frac{2\varepsilon}{\alpha-1}=
 -\frac{2}{\alpha-1}
 \left.\left(1-\frac{\delta_x}{d}\right)\right|_{\alpha=\alpha{}_{\log}}
   = \delta_x-d.
\end{equation}
In agreement to dimensional considerations, the renormalized
Green function has the form
\[
G_R(t,r)=(\nu t)^{-d/2}\chi\left(\frac{r^2}{\nu t},\varpi \nu
t,\frac 1{t\mu^2\nu} \right), \quad r=|\mathbf{x}-\mathbf{x}'|,
\]
where $\chi$ is a scaling function of the dimensionless variables. The
dependence on $\eta$ is not displayed explicitly, because the
derivatives with respect to this parameter do not enter into the
scaling equation. It follows from (\ref{sceq}) that at the fixed
point $\chi$ satisfies  the equation
\[
\left[z\frac{\partial}{\partial
z}-\frac{\gamma_*}{2}\right]\chi(s,y,z)=0,
\]
its general solution is $\chi(s,y,z)=s^{\gamma_*/2}\psi(s,y)$
where $\psi$ is an arbitrary function of the first and second
arguments. For the Green function (\ref{Green}) we then obtain
\begin{equation}\label{gras}
  G(t,r)\sim t^{-d/2-\gamma_*/2}\psi\left(\frac{r^2}{\nu t},\varpi \nu
t\right)=t^{-\delta_x/2}\psi\left(\frac{r^2}{\nu t},\varpi \nu
t\right),\quad \delta_x>d,
\end{equation}
where the form of the scaling function $\psi$ is not determined by
the equation (\ref{sceq}). Although the value of $\gamma_*$ in
(\ref{gan}) and the solution (\ref{gras}) have been obtained
without calculation of the constant $Z$, such a calculation is
necessary to check the existence and stability of the fixed point.
Within the $\varepsilon-$expansion these facts can be verified
already in the  one-loop order. We perform it in the forthcoming
section.

We conclude the section with a remark on the arguments of scaling
function $\psi$  for the planar 3-regular graphs of order $2N$ with the standard
orientation (i.e. the $2N$-honeycomb). In
the Sec.~2, we have mentioned that they are equivalent to the
sphere of radius $\rho$. The surface area of the sphere equals
to  $2\pi N =4\pi \rho^2,$ and therefore $\rho=\sqrt{N/2}.$ The
relevant Gilkey coefficients and the asymptotic behavior of the Feynman
propagator as $t\to 0$ are given at the end of Appendix A. One can
see that the corrections to the standard diffusion kernel risen by
the space curvature can be naturally interpreted as the
corrections due to the finite size of the 3-regular subgraph.
Then, the scaling function $\psi$ is
\[
\psi=\psi\left(\frac{r^2}{\nu t},\frac{2\nu t}{N}\right),
\]
and it can be calculated by means of diagram expansion. In the
thermodynamic limit, $N\to \infty,$ (when the the graph $\Gamma$
is large and regular) these corrections vanish and the scaling
function depends only on one argument $r^2/\nu t$ relevant to the flat
space. The result on the critical scaling (\ref{gras}) is still valid in the
 thermodynamic limit.

\section{A fixed point of RG transformation and heat-kernel expansion}
\noindent

The main technical difficulty concerning diagram computations in curved
spacetime is the lack of a global momentum-frequency
representation. The momentum space is associated to each point
$x'$ by the Fourier transformation
\[
f(x,x')=\int \frac{d^dx}{(2\pi)^d} e^{iky} F(k;x')
\]
where $ky=g^{ij}k_iy_j$ which does not help much in calculations.
In the flat metric, the calculation of the residue $c_1$ in (\ref{Z}),
in the one-loop order, can be performed in the standard way using
the momentum-frequency representation \cite{AH},
\begin{equation}\label{c1flat}
 \left. c_1\right|_{\mathrm{flat}}= \frac{\eta\nu}{4\pi}
\alpha^{-1/(\alpha-1)}, \quad Z=1+\frac{c_1}{\varepsilon}.
\end{equation}
It corresponds to the fixed point
\begin{equation}\label{fp}
  \left. \eta_*\right|_{\mathrm{flat}}= \frac{\varepsilon}{\nu(\alpha-1)}
\end{equation}
which is positive and stable for small  $\varepsilon>0$ and
$\alpha>1$. In the curved space the new singularities would arise
at small scales (as $\mathbf{x}\to\mathbf{x}'$) even for the
metrics of constant curvature. They do not affect the
renormalization group procedure (at least at the one-loop order)
at large scales, but breaks down the perturbation theory for the
even space dimensions.

The necessary regularization of effective action  can be performed
by using the heat-kernel expansion which does not depend on the
spacetime topology (see, for instance, \cite{T82}).
  We consider the operator $Q_2$ with the kernel
\[
Q_2 =-\frac{\delta^2 \mathcal{S_R}}{\delta U {\ }\delta V}, \quad
U,V = u, u',
\]
where $\mathcal{S}_R$ is the renormalized action functional of the
model we consider. The one-loop amendment to the renormalized
action is given by
\begin{equation}\label{L1}
  L_1=-\frac 12 \mathrm{Tr} \ln
\left(Q_2\mathbf{\Delta}^{-1}\right)
\end{equation}
where the propagator matrix is given by
\[
\mathbf{\Delta}=\left(
\begin{array}{cc}
0 & \Delta_{uu'} \\
  \Delta_{u'u} & 0
\end{array}
\right).
\]
The counterterm corresponding to the only divergent diagram block
is linear in the auxiliary field $u'$. Variation of $L_1$ with
respect to the auxiliary field
\[
\delta_{u'} L_1=\frac 12 \mathrm{Tr}\left( \mathbf{\Delta}Q^{-1}_2
\delta_{u'}(Q_2\mathbf{\Delta}^{-1})\right), \quad
\delta_{u'}(Q_2\mathbf{\Delta}^{-1})= -\alpha(\alpha-1)\eta\nu
u^{\alpha-2}\mathbf{\Delta}^{-1},
\]
can be written as
\[
\delta_{u'}L_1=-\frac 12\delta_{u'}\left\{ \mathrm{Tr}
\int_0^{\infty}\frac{d\tau}{\tau}\exp (-\tau\cdot
Q_2\mathbf{\Delta}^{-1}) \right\}.
\]
Let us suppose that
$\Lambda(t,\mathbf{x},\mathbf{y},Q_2{\mathbf{\Delta}}^{-1})$ is the
kernel associated with the operator $\exp(-\tau\cdot
Q_2{\mathbf{\Delta}}^{-1})$. Then, the one-loop contribution
(\ref{L1}) reads as
\begin{equation}\label{rtyu}
  L_1=-\frac12 \int dv_x\int^\infty_0 \frac {d\tau}{\tau}
\Lambda(\tau,\mathbf{x},\mathbf{x},Q_2{\mathbf{\Delta}}^{-1}).
\end{equation}
Since the divergences are purely logarithmic, one can neglect the
inhomogeneity of the scalar field $u$ in $\mathbf{x}$ while
calculating the relevant counterterm. Being treated as a constant,
the scalar $u$ renormalizes the scalar curvature,
\[
R\to R-\frac{\alpha\nu\eta}{\xi}u^{\alpha-1},
\]
so that one can define the modified Feynman propagator,
$\Delta_{uu'}\to\Delta_{uu'}'$.

Then, it appears that the kernel
$\Lambda(\tau,\mathbf{x},\mathbf{x},(\mathbf{\Delta}')^{-1})$
meets the "heat equation"
\[
[\partial_ \tau+(\mathbf{\Delta}')^{-1}]
\Lambda(\tau,\mathbf{x},\mathbf{x},(\mathbf{\Delta}')^{-1})=0
\]
and therefore allows for the asymptotic expansion as $\tau\to 0+$,
in the limit $\mathbf{x}\to\mathbf{x}',$
\begin{equation}\label{r4t}
 \Lambda(\tau,\mathbf{x},\mathbf{x},Q_2{\mathbf{\Delta}}^{-1})\sim
(4\pi\tau)^{-d/2}\sum_{m=0}^{\infty}\tau^m
E_m(\mathbf{x},\mathbf{\Delta}'),
\end{equation}
in which $E_m(\mathbf{x},\mathbf{\Delta}')$ are the relevant
Gilkey coefficients \cite{Gilk1}-\cite{Gilk4}. In the case of a constant curvature metric,
the first coefficients can be found in the Appendix A
(\ref{Gilkey}). Substituting the expansion (\ref{r4t}) into the
integral representation (\ref{rtyu}), one can perform the
integrations with respect to the parameter $\tau$ with a cut off
$\tau_0$. As a result, one arrives at the power series,
\begin{equation}\label{fg}
-  \frac 1{(4\pi)^{d/2}}\sum_{m=0}^{\infty}
\frac{\tau_0^{m-d/2}}{m-d/2}E_m(\mathbf{x},\mathbf{\Delta}'),
\end{equation}
being singular for integer $m$ such that $d=2m$. It is obvious
that the breakdowns of perturbation theory at small scales would
happen for the even space dimensions. Extracting the pole part of
$L_1$ one finds that
\begin{equation}\label{pole}
P.P.[L_1]=\frac 1{(\delta_x-d)}\frac 1{(4\pi)^{d/2}}\int dv_x
E_{d/2}(x)=\frac 1{\varepsilon}\frac {\delta_x}{(4\pi)^{d/2}}\int
dv_x E_{d/2}(x)
\end{equation}
where $E_{d/2}$ are the relevant Gilkey coefficient provided that
$d$ is even. It is important to mention that in (\ref{pole}) the
singularity coincides with that of (\ref{c1flat}). The
residue in (\ref{pole}) has the "good" signature and does not
break the stability of the fixed point of RG transformation.

\section{Discussion and Conclusion}
\noindent

In the present paper, we have studied the transport through the large
complex networks containing regular subgraphs. We have considered
such networks as  generalized trees in which two types of nodes
are allowed: simple nodes and supernodes. We have supposed that
the supernodes are either the subgraphs with many polygons or
$k$-regular subgraphs. In particular, we have discussed the case of
3-regular subgraphs which can be treated as the complete Riemann
curved surfaces characterized by  finite areas and can be compactified.

The diffusion process taking place on such a complex network is considered
as a generalized Brownian
motion with arbitrary boundary conditions. Its random dynamics is then
described by a functional integral.
 We have studied  the long time large scale
asymptotic behavior of the Green function for the nonlinear
diffusion equation defined on the complex network supplied with an
integrable initial condition. The transport trough the complex
network is of a strongly nonlinear nature being affected by both
the varying effective space dimension between supernodes and the
space curvature within a supernode.

In the regular graphs endowed with the standard orientation of
edges (the cyclic ordering of edges to be the same for all
nodes), the effect of curvature within the supernodes reveals
itself by the finite size corrections to the scaling behavior. In
the space of even dimensions , the curvature results in the
additional singularities of Green function at small scales and
calls for the special regularization. We have to stress that in
contrast to the case of gravitational field, we are not restricted
by the equivalence principle while considering models of complex
networks. The main technical difficulty of computations in the
curved space metrics is the lack of the global frequency- momentum
representation. We have used the heat-kernel expansion which does
not depend on the space time topology to regularize the
effective action at small scales in the one-loop order.

Our approach to the complex networks containing large
well-structured regular subgraphs could be used in studies of
spreading of viruses \cite{VVB} or social epidemics
\cite{BlKrurg}, traffic properties and the modelling of various
ecological webs.

\section{Acknowledgements}
\noindent

One of the authors (D.V.)  benefits from a scholarship of the
Alexander von Humboldt Foundation (Germany) and of a support of
the DFG-International Graduate School "Stochastic and real-world
problems" that he gratefully acknowledges.

\vspace{1cm}

\textbf{Appendix A. Feynman propagator in curved space}

\vspace{0.5cm}

We present the explicit solution of the linearized diffusion
problem, in the $d$-dimensional curved space metric, for
 $G(\mathbf{x},\mathbf{x}',t)$ satisfying
$\lim_{\mathbf{x}'\to
0}G(\mathbf{x},\mathbf{x}',t)=G(\mathbf{x},t)$ and $\lim_{t\to
0}G(\mathbf{x},\mathbf{x}',t)=\delta(\mathbf{x}-\mathbf{x}')$. The
details can be found in \cite{BuPa} for field theories and in
\cite{Bala} (see also references therein) for classical models.

The general solution is
\begin{equation}\label{prop}
\Delta_{uu'}(\mathbf{x},\mathbf{x}',t)=\theta(t)\frac{\exp(-\xi_0\eta_0Rt)}{(4\pi
t)^{d/2}}\exp\left(-\frac{\sigma(\mathbf{x},\mathbf{x}')}{2t}\right)
\Delta^{1/2}(\mathbf{x},\mathbf{x}')\Omega(\mathbf{x},\mathbf{x}',t)
\end{equation}
where $\theta(t)$ is the Heaviside function,
$\sigma(\mathbf{x},\mathbf{x}')$ equals to half the square of the
geodesic distance between $\mathbf{x}$ and $\mathbf{x}'$, and
$\Delta(\mathbf{x},\mathbf{x}')$ is the van Vleck determinant,
\[
\Delta(\mathbf{x},\mathbf{x}')=-\frac{\det\left(\partial_i\partial_j
\sigma(\mathbf{x},\mathbf{x}')\right)}
{\sqrt{g(\mathbf{x})g(\mathbf{x}')}},
\]
which reduces to unity in flat space, $R$ is the scalar curvature.
The function $\Omega(\mathbf{x},\mathbf{x}',t)$ allows for the
following series expansion in the limit
$\mathbf{x}'\to\mathbf{x}$:
\[
\lim_{\mathbf{x}'\to\mathbf{x}}
\Omega(\mathbf{x},\mathbf{x}',t)=1+
\sum_{l=1}^{\infty}t^lE_l(\mathbf{x})
\]
valid in the limit $t\to 0$ where $E_l(\mathbf{x})$ are known in
the literature as Gilkey coefficients \cite{Gilk1}-\cite{Gilk4}. For the diffusion equation,
in the flat metric, the only coefficient which contributes is
$E_0$ and we recover the well known standard diffusion kernel.

The planar 3-regular graph of order $2N$ with the standard orientation (the
cyclic ordering of edges is taken the same for each node), a $2N$-honeycomb,
 corresponds to a sphere
of radius $\rho =\sqrt{N/2}.$ The Ricci scalar curvature is then
\[
R=\frac{2}{\rho^2}=\frac 4{N},
\]
and the Gaussian curvature equals to $\kappa=2/N.$ The Gilkey
coefficients reduces to
\[
E_0=1, \quad E_1=\frac 2{3N},
\]
\begin{equation}\label{Gilkey}
 E_2=\frac{4}{15}\frac 1{N^2}, \quad E_3=\frac{32}{315}\frac
 1{N^3}.
\end{equation}
Then, in the limit $\mathbf{x}'\to\mathbf{x}$ and $t\to 0$, the
Feynman propagator exhibits the following dependence  on the
size of 3-regular subgraph:
\begin{equation}
\label{last}
  \Delta_{uu'}(\mathbf{x},t)=\theta(t)\frac{\exp\left(-\frac{x^2}{4\nu_0
t}\right)}{4\pi\nu_0 t} \left( 1+\frac
13\left(\frac{2\nu_0t}{N}\right)+\frac
1{15}\left(\frac{2\nu_0t}{N}\right)^2 +
\frac{4}{315}\left(\frac{2\nu_0t}{N}\right)^3 +\ldots\right).
\end{equation}

\begin{figure}[ht]
\noindent
%\begin{minipage}[b]{.36\linewidth}
\begin{center}
\epsfig{file=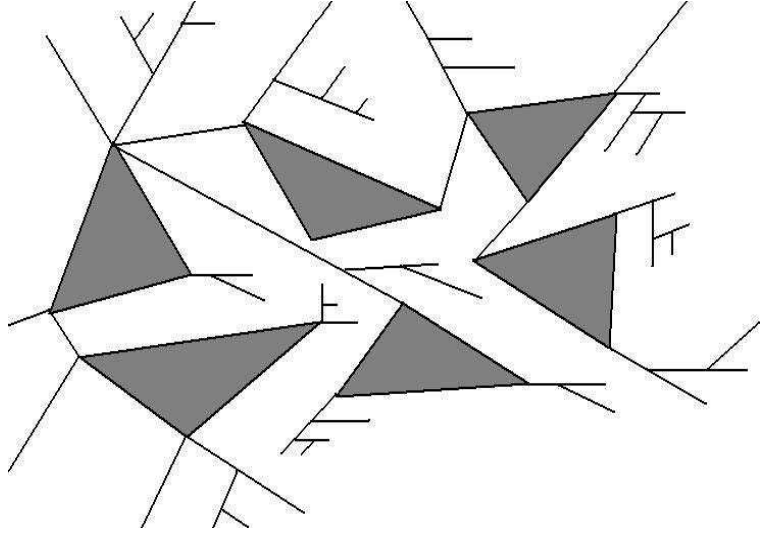,width=10cm, height=7 cm}
\end{center}
%\end{minipage}
\caption{In the model, we consider the complex networks as the
generalized trees with two types of nodes: the simple nodes
(probably of low connectivity) and the supernodes which are either
the subgraphs with many triangles or the $k$-regular subgraphs.}
\end{figure}

\begin{figure}[ht]
\noindent
%\begin{minipage}[b]{.36\linewidth}
\begin{center}
\epsfig{file=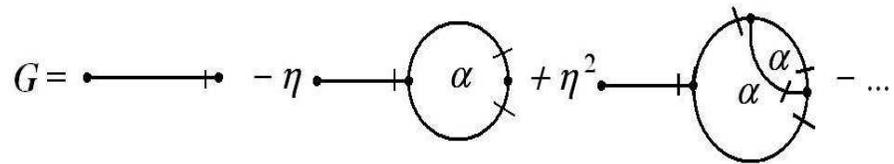,width=12cm, height=2.5 cm}
\end{center}
%\end{minipage}
\caption{Three diagrams of the Feynman diagram expansion for the
Green function of the nonlinear diffusion equation.}
\end{figure}

\end{document}